\documentclass{wscpaperproc}
\usepackage{latexsym}
\usepackage{graphicx}
\usepackage{mathptmx}
\usepackage{placeins}

\usepackage{amsmath}
\usepackage{amsfonts}
\usepackage{amssymb}
\usepackage{amsbsy}
\usepackage{amsthm}
\usepackage[pdftex,colorlinks=true,urlcolor=blue,citecolor=black,anchorcolor=black,linkcolor=black]{hyperref}

\newtheoremstyle{wsc}% hnamei
{3pt}% hSpace abovei
{3pt}% hSpace belowi
{}% hBody fonti
{}% hIndent amounti1
{\bf}% hTheorem head fontbf
{}% hPunctuation after theorem headi
{.5em}% hSpace after theorem headi2
{}% hTheorem head spec (can be left empty, meaning `normal')i

\theoremstyle{wsc}

\begin{document}

%***************************************************************************
% AUTHOR: AUTHOR NAMES GO HERE
% FORMAT AUTHORS NAMES Like: Author1, Author2 and Author3 (last names)
%
%		You need to change the author listing below!
%               Please list ALL authors using last name only, separate by a comma except
%               for the last author, separate with "and"
%

% setting up general page style
\pagestyle{fancyplain}

% setting up page style of first page
\thispagestyle{plain}
\firstPageHead{}

% setting up running header (authors) of subsequent pages
\chead{\fancyplain{}{\itshape Mohammadrezaei, Giovannelli, Lane, and Gra{\v{c}}anin}}

% setting up seperation parameters
%\headsep=72pt
\rhead{}
\cfoot{}
\renewcommand{\headrulewidth}{0pt} % (renewcommand needed in fancyhdr to remove top decorative line)
%\headrulewidth=0pt  % ("setlength" needed in fancyheading to remove top decorative line)

%%%%%%%%%%%%%%%%%%%%%%%%%%%%%%%%%%%%%%%%%%%%%%%%%%%%%%%%%%%%%%%%%%%%%%%%%%%%%%
%                                                                            %
%     THESE COMMANDS ARE REQUIRED TO WORK WITH WSC.BST TO MAKE BIBLIO     %
%                                                                            %
%%%%%%%%%%%%%%%%%%%%%%%%%%%%%%%%%%%%%%%%%%%%%%%%%%%%%%%%%%%%%%%%%%%%%%%%%%%%%%
\makeatletter
\let\@internalcite\cite
\def\cite{\def\@citeseppen{-1000}%
    \def\@cite##1##2{(##1\if@tempswa , ##2\fi)}%
    \def\citeauthoryear##1##2##3{##1 ##3}\@internalcite}
\def\citeNP{\def\@citeseppen{-1000}%
    \def\@cite##1##2{##1\if@tempswa , ##2\fi}%
    \def\citeauthoryear##1##2##3{##1 ##3}\@internalcite}
\def\citeN{\def\@citeseppen{-1000}%
%  Pierre L'Ecuyer's fix for multiple cite bug
%  Added by Paul J Sanchez on 4 October 2001
%   \def\@cite##1##2{##1\if@tempswa , ##2)\else{)}\fi}%
%   \def\citeauthoryear##1##2##3{##1 (##3}\@citedata}
    \def\@cite##1##2{##1\if@tempswa, ##2)\else{}\fi}%
    \def\citeauthoryear##1##2##3{##1 (##3)}\@citedata}
\def\citeA{\def\@citeseppen{-1000}%
    \def\@cite##1##2{(##1\if@tempswa , ##2\fi)}%
    \def\citeauthoryear##1##2##3{##1}\@internalcite}
\def\citeANP{\def\@citeseppen{-1000}%
    \def\@cite##1##2{##1\if@tempswa , ##2\fi}%
    \def\citeauthoryear##1##2##3{##1}\@internalcite}
\def\shortcite{\def\@citeseppen{-1000}%
    \def\@cite##1##2{(##1\if@tempswa , ##2\fi)}%
    \def\citeauthoryear##1##2##3{##2 ##3}\@internalcite}
\def\shortciteNP{\def\@citeseppen{-1000}%
    \def\@cite##1##2{##1\if@tempswa , ##2\fi}%
    \def\citeauthoryear##1##2##3{##2 ##3}\@internalcite}
\def\shortciteN{\def\@citeseppen{-1000}%
%  Pierre L'Ecuyer's fix for multiple cite bug
%  Added by Paul J Sanchez on 2 September 2002
%  should have caught this last year...
%   \def\@cite##1##2{##1\if@tempswa , ##2)\else{)}\fi}%
%   \def\citeauthoryear##1##2##3{##2 (##3}\@citedata}
% Shane G. Henderson fix for extra right bracket at end of optional material June 8, 2005
%    \def\@cite##1##2{##1\if@tempswa, ##2)\else{}\fi}%
    \def\@cite##1##2{##1\if@tempswa, ##2\else{}\fi}%
    \def\citeauthoryear##1##2##3{##2 (##3)}\@citedata}
\def\shortciteA{\def\@citeseppen{-1000}%
    \def\@cite##1##2{(##1\if@tempswa , ##2\fi)}%
    \def\citeauthoryear##1##2##3{##2}\@internalcite}
\def\shortciteANP{\def\@citeseppen{-1000}%
    \def\@cite##1##2{##1\if@tempswa , ##2\fi}%
    \def\citeauthoryear##1##2##3{##2}\@internalcite}
\def\citeyear{\def\@citeseppen{-1000}%
    \def\@cite##1##2{(##1\if@tempswa , ##2\fi)}%
    \def\citeauthoryear##1##2##3{##3}\@citedata}
\def\citeyearNP{\def\@citeseppen{-1000}%
    \def\@cite##1##2{##1\if@tempswa , ##2\fi}%
    \def\citeauthoryear##1##2##3{##3}\@citedata}
%
% \@citedata and \@citedatax:
%
% Place commas in-between citations in the same \citeyear, \citeyearNP,
% \citeN, or \shortciteN command.
% Use something like \citeN{ref1,ref2,ref3} and \citeN{ref4} for a list.
%
\def\@citedata{%
    \@ifnextchar [{\@tempswatrue\@citedatax}%
                  {\@tempswafalse\@citedatax[]}%
}

\def\@citedatax[#1]#2{%
\if@filesw\immediate\write\@auxout{\string\citation{#2}}\fi%
  \def\@citea{}\@cite{\@for\@citeb:=#2\do%
    {\@citea\def\@citea{, }\@ifundefined% by Young
       {b@\@citeb}{{\bf ?}%
       \@warning{Citation `\@citeb' on page \thepage \space undefined}}%
{\csname b@\@citeb\endcsname}}}{#1}}%

% don't box citations, separate with ; and a space
% also, make the penalty between citations negative: a good place to break.
%
\def\@citex[#1]#2{%
\if@filesw\immediate\write\@auxout{\string\citation{#2}}\fi%
  \def\@citea{}\@cite{\@for\@citeb:=#2\do%
    {\@citea\def\@citea{; }\@ifundefined% by Young
       {b@\@citeb}{{\bf ?}%
       \@warning{Citation `\@citeb' on page \thepage \space undefined}}%
{\csname b@\@citeb\endcsname}}}{#1}}%

% (from apalike.sty)
% No labels in the bibliography.
%
\def\@biblabel#1{}
\makeatother

%\newlength{\bibhang}
%\setlength{\bibhang}{2em}

% Indent second and subsequent lines of bibliographic entries. Taken
% from openbib.sty: \newblock is set to {}.
% \renewcommand{\refname}{REFERENCES}

\newdimen\bibindent
\bibindent=0.0em
% SEC: was \def\thebibliography#1{\section*{\refname\@mkboth
% SEC: was   {\uppercase{\refname}}{\uppercase{\refname}}}\list
\def\thebibliography#1{\section*{\refname}\list
   {}{\settowidth\labelwidth{[#1]}
   \leftmargin\parindent
   \itemindent -\parindent
   \listparindent \itemindent
   \itemsep 0pt
   \parsep 0pt}
   \def\newblock{}
   \sloppy
   \sfcode`\.=1000\relax}

           % Set up BiBTeX macros

% needed to make the tex document look more like the word counterpart :-(
\setlength{\baselineskip}{12.7pt}

% AUTHOR: Enter the title, all letters in upper case
\title{A DIGITAL TWIN BASED APPROACH TO SMART LIGHTING DESIGN}

% AUTHOR: Enter the authors of the article, see end of the example document for further examples
\author{Elham Mohammadrezaei\\
Alexander Giovannelli\\
Logan Lane\\
Denis Gra{\v{c}}anin\\[12pt]
Department of Computer Science\\
Virginia Tech\\
2202 Kraft Drive\\
Blacksburg, VA 24060, USA\\
}

\maketitle

\section*{ABSTRACT}

Lighting has a critical impact on user mood and behavior, especially in architectural settings.
%Smart home technology and LED lighting are changing the face of home illumination.
Consequently, smart lighting design is a rapidly growing research area. % that explores the use of new lighting functionality for architectural spaces.
%Therefore, it is essential to develop methods and techniques to support smart lighting design.
We describe a digital twin-based approach to smart lighting design that uses an immersive virtual reality digital twin equivalent (virtual environment) of the real-world, physical architectural space to explore the visual impact of light configurations.
The CLIP neural network is used to obtain a similarity measure between a  photo of the physical space with the corresponding rendering in the virtual environment.
A case study was used to evaluate the proposed design process.
The obtained similarity value of over 87\% demonstrates the utility of the proposed approach.

\section{INTRODUCTION}
\label{sec:intro}

\begin{figure}[!b]
\centering   
\includegraphics[width=1\textwidth]{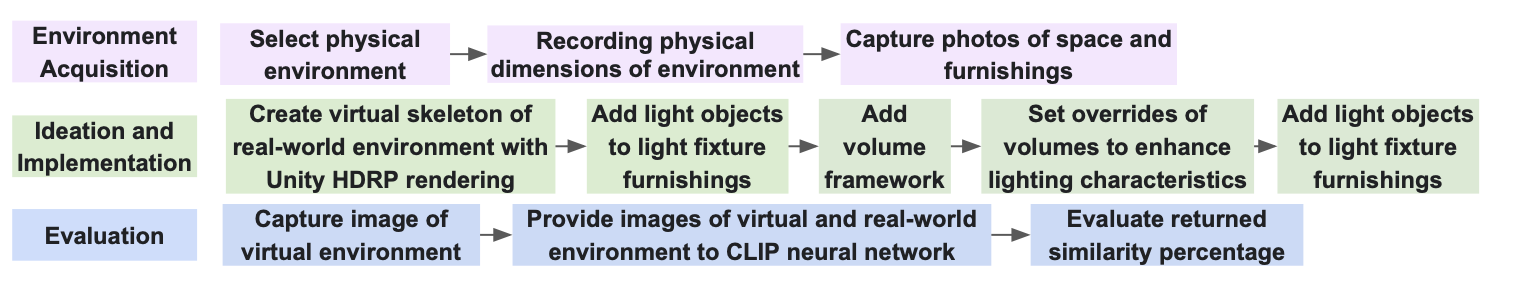}
% \caption{Rendering: room with no volumes to reflect light.}
\caption{A diagram of our digital twin system approach.}
\label{fig:approach}
\end{figure}

As smart lighting technology progresses, the ability to simulate and visualize its potential impact within environments prior to  construction is increasingly important.
The primary goal of our research is establishing an immersive virtual reality (VR) system to simulate a given physical environment, empowering near-identical smart lighting design within a digital twin~\cite{Mohammadrezaei-2022-a}.
The system supports real-time changes to lighting fixture settings such as brightness, color, and temporality, such that users are able to interact with the lighting system in the digital twin and witness the impact of those changes to lighting settings immediately. We describe a systematic approach that users can follow to create their own digital twin of a real-world, physical environment (Figure~\ref{fig:approach}).
(Figure~\ref{fig:approach}).
%This approach aims to resolve the gaps in the capabilities of physical and hybrid-based light setting systems by directing users toward commercially accessible hardware and freely-distributed software technologies.
The users are guided on how to use hardware devices to capture details of the real-world environment for which they would like to produce a digital twin.
These details include lighting values of bulbs, furnishing, and dimensional measurements of the environment and its contents. We use real-time lighting  %in the digital twin's creation, such that users will be able to see 
to show the impact of changes to brightness, color and temporality immediately, instead of a static, `baked' environment that requires re-rendering when a lighting value is updated.
This provides the existing benefits of physical and hybrid-based lighting so that users receive immediate feedback on lighting setting changes on their digital twin the same as the real-world equivalent.
%By following the guidelines derived from our own system's development using these technologies and design requirements, we empower users to produce their own digital twins of real-world environments with homogeneous lighting fixture values.
%We now describe these technologies and the overall digital twin system design.

\section{RESULTS}

To validate our digital twin's similarity to our photo capturing the real-world office, we generated three  renderings: one with missing furniture and incorrect lighting color via the emission temperature setting, one with no local volumes to reflect, refract, blend light and create shadows, and the final one with optimal volume, furniture, and lighting values.
As we enhanced the scene to get the digital twin's lighting to match the lighting of the real environment, the returned similarity value increased (Table~\ref{tab:SimilarityPercentages}). 
\begin{figure}[!ht]
\centering
\includegraphics[height=2.3in]{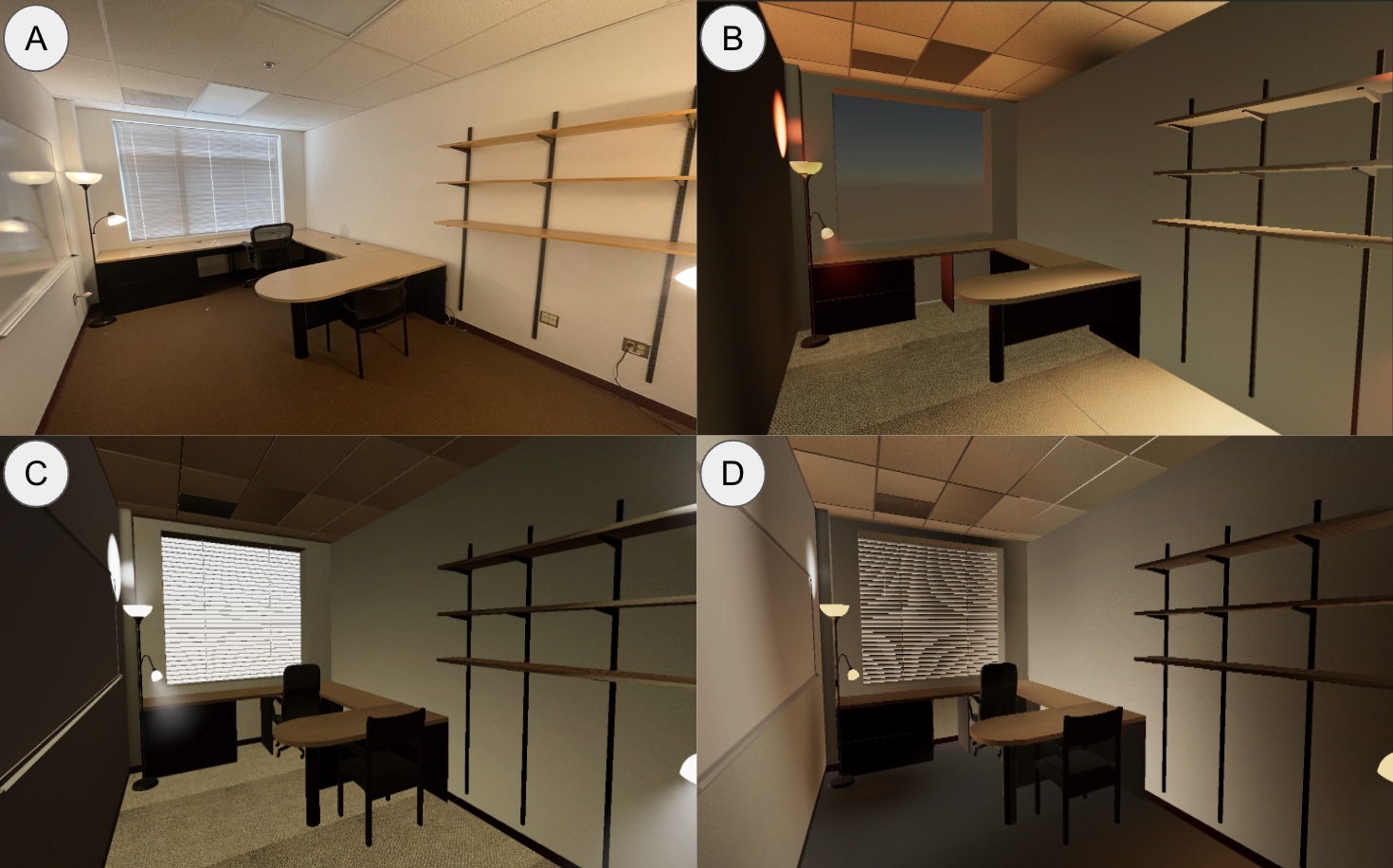}
% \hspace*{0.6in}\includegraphics[height=1.25in]{REAL.jpg}
% \includegraphics[height=1.25in]{lower_temp.jpg}
% \includegraphics[height=1.25in]{bad_1.jpg}
% \includegraphics[height=1.25in]{4.jpg}
% \caption{Photo: the real-world, physical office space.}
\caption{\textbf{A:} Photo of the real-world, physical office space.
\textbf{B:} Rendering with missing furniture and incorrect lighting values.
\textbf{C:} Rendering with no volumes to reflect light.
\textbf{D:} Rendering with optimal volume, furniture, and lighting values.
}
\label{fig:twin}
\end{figure}
%\FloatBarrier

%\begin{figure}[!ht]
%\centering
%\includegraphics[width=0.75\textwidth]{4.jpg}
% \caption{Rendering: optimal volume, furniture, and lighting values.}
%\caption{Rendering with optimal volume, furniture, and lighting values.}
%\label{fig:Twin}
%\end{figure}

% \begin{figure}[!ht]
% \centering   
% \includegraphics[width=0.75\textwidth]{bad_1.jpg}
% % \caption{Rendering: room with no volumes to reflect light.}
% \caption{Rendering with no volumes to reflect light.}
% \label{fig:NoVolume}
% \end{figure}

% \begin{figure}[!ht]
% \centering
% \includegraphics[width=0.75\textwidth]{lower_temp.jpg}
% % \caption{Rendering: room with missing furniture and incorrect lighting values.}
% \caption{Rendering with missing furniture and incorrect lighting values.}
% \label{fig:WrongLighting}
% \end{figure}

%From these three unique images, we found that lighting played a significant role in altering the similarity value returned by the CLIP neural network. 

\begin{table}[!ht]
\centering
\caption{Digital twin similarity scores per lighting condition change.}
{\small
\begin{tabular}{|c|l|c|}
\hline
\textbf{Figure~\ref{fig:twin}} & \multicolumn{1}{c|}{\textbf{Description}} & \textbf{CLIP Similarity Percentage} \\ \hline
B & Incorrect lighting and missing furniture & 77.168 \\ \hline
C & Missing Volume framework & 78.427 \\ \hline
D & Final Digital Twin & 87.665 \\ \hline
\end{tabular}
}
\label{tab:SimilarityPercentages}
\end{table}

\section{CONCLUSION}

We described a digital twin-based approach to smart lighting design that provides designers with a way to evaluate the visual impact of light sources and fixture configurations in architectural settings.
Users can use head-mounted displays to fully immerse themselves in a virtual environment, explore different light designs and accurately evaluate the visual outcomes.
%With the continual advancement of smart lighting design, the need to establish such an approach to simulate and visualize light designs grows as well.
%Smart lighting has the ability to influence the environment’s mood, increase the energy efficiency of lighting fixtures, and spotlight locations according to pedestrian presence.
We believe our case provides a novel contribution to this rapidly growing research area.

%\nocite{*}

\footnotesize

\bibliographystyle{wsc}

% AUTHOR: Include your bib file here
%\bibliography{references}
\bibliography{wsc_2022_paper_lighting}

\end{document}